\documentclass[a4paper]{article}

\usepackage{lmodern}
\usepackage[T1]{fontenc}
\usepackage[utf8]{inputenc}
\usepackage{amssymb,amsmath}
\usepackage{graphicx}
\usepackage{layouts}

\usepackage{hyperref}
\hypersetup{unicode=true,
            pdfborder={0 0 0},
            breaklinks=true}
\usepackage[gen]{eurosym}
\usepackage{longtable,booktabs}
\IfFileExists{parskip.sty}{%
\usepackage{parskip}
}{
\setlength{\parindent}{0pt}
\setlength{\parskip}{6pt plus 2pt minus 1pt}
}
\newcommand{\rfig}[1]{Figure~\ref{figure_#1}}

\newcommand{\basicfig}[3]{%
      \begin{figure}[tbp]%
      \centering%
              #1
      \caption{#3}
      \label{figure_#2}
      \end{figure}
}

\newcommand{\boxfig}[3]{\basicfig{\mbox{#1}}{#2}{#3}}

\newcommand{\nfig}[4][]{ 
      \boxfig{\includegraphics[#1]{#2}}{#3}{#4}
}
\newcommand{\fig}[3][]{\nfig[#1]{#2}{#2}{#3}}

\title{Flat-field and colour correction for the Raspberry Pi camera module}
\usepackage{authblk} 
\author[1*]{Richard Bowman}
\author[2]{Boyko Vodenicharski} 
\author[1]{Joel Collins} 
\author[1]{Julian Stirling} 
\affil[1]{Department of Physics, University of Bath} 
\affil[2]{Cavendish Laboratory, University of Cambridge}
\affil[*]{r.w.bowman@bath.ac.uk}

\begin{document}

\maketitle

\begin{abstract}

The Raspberry Pi camera module is widely used in open source hardware projects as a low cost camera sensor. However, when the stock lens is removed and replaced with other custom optics the sensor will return a non-uniform background and colour response which hampers the use of this excellent and popular image sensor. This effect is found to be due to the sensor's optical design as well as due to built-in corrections in the GPU firmware, which is optimised for a short focal length lens. In this work we characterise and correct the vignetting and colour crosstalk found in the Raspberry Pi camera module v2,  presenting two measures that greatly improve the quality of images using custom optics. First, we use a custom ``lens shading table'' to correct for vignetting of the image, which can be done in real time in the camera's existing processing pipeline (i.e. the camera's low-latency preview is corrected).  The second correction is a colour unmixing matrix, which enables us to reverse the loss in saturation at the edge of the image, though this requires post-processing of the image.  With both of these corrections in place, it is possible to obtain uniformly colour-corrected images, at the expense of slightly increased noise at the edges of the image.
\end{abstract}

\begin{longtable}[]{@{}l@{}}
\begin{minipage}[t]{0.97\columnwidth}\raggedright\strut

\subsection{Metadata Overview}\label{h.akaipbqoqfs8}

Main design files: \url{https://gitlab.com/bath_open_instrumentation_group/picamera_cra_compensation}

Target group: microscopists, optical instrument builders, people using custom imaging systems

Skills required: Python programming - easy/moderate, Arduino - easy, 3D printing - easy;

Replication\textsuperscript{\protect\hyperlink{cmnt1}{{[}a{]}}}{:
}{\protect\url{https://gitlab.com/bath_open_instrumentation_group/picamera_cra_compensation}}

See section ``Build Details'' for more detail.

\subsection{Keywords}\label{h.kdz351yp7g7c}

{(required)}{~imaging; image processing; open source; Python;}

\strut\end{minipage}\tabularnewline
\bottomrule
\end{longtable}

\section{Introduction}
The Raspberry Pi single-board computer\cite{raspberry_pi_foundation} and its accompanying camera module are staple components of many open hardware projects\cite{maiachagas2017,nunez2017,sharkey2016}.  The ability to embed a small and inexpensive but capable computer enables real-time display and processing of images, and is an excellent basis for scientific instruments that require an imaging component.  The Raspberry Pi Camera Module is now in its second version, and is a small ($24\times25\,$mm) breakout board that connects a Sony IMX219 image sensor \cite{imx219_datasheet} to the Raspberry Pi via a MIPI ribbon cable.  This is supported by real-time code running in the GPU firmware, that makes it possible to display a high resolution, low-latency video preview on the Raspberry Pi, and to capture compressed and raw images and video sequences.  Popular and well-documented APIs exist in C and Python \cite{picamera} to control the camera, and there is a strong community of users working with it.  There are a wide range of microscopy projects making use of the Raspberry Pi camera including the OpenFlexure Microscope \cite{sharkey2016}, FlyPi \cite{maiachagas2017}, a fluorescence imaging system \cite{nunez2017}, and various others \cite{coward2017, publiclab_community_microscope, grant2019, li2019}.  There have been efforts made in the past to correct the camera for uniform, flat-field response \cite{pagnutti2017}.  This manuscript adds real-time flat-field response (referred to as the ``lens shading table'') and an analysis of colour crosstalk that enables uniform colour response across the sensor.

\subsection{Chief Ray Angle compensation}
The current camera module (version 2) uses the Sony IMX219 camera module - this is an 8 megapixel back-illuminated CMOS image sensor, intended for use in mobile phones and tablets \cite{imx219_datasheet}.  It is supplied with a $3.04\,$mm focal length lens, and the sensor measures $4.6\,$mm across the diagonal which provides the wide field of view typically expected of a mobile phone camera \cite{picamera_hardware_docs}.  The low light performance of the IMX219 is improved by two key features - firstly, the sensor is ``back illuminated'', meaning that the light enters the sensor and hits the light-sensitive region without having to pass through any of the readout electronics - this enables a greater fill fraction and consequently improves efficiency.  Secondly, a lenslet array is bonded to the image sensor chip.  This concentrates the light onto each pixel, further increasing the fraction of the available light that is collected \cite{imx219_datasheet}.

Each pixel is positioned underneath one lenslet, such that light incident on that lenslet (which may be assumed to be roughly collimated) is focused onto the light-sensitive area of the pixel.  This is a relatively simple picture for pixels in the centre of the sensor, where the light is normally incident on the sensor.  However, at the edges of the sensor, the stock lens (which is typical of mobile phone camera lenses) means that light is not normally incident, and is travelling at an angle.  If the lenslets at the edge of the sensor were still directly above the light-sensitive regions of the pixels, the light would not focus in the correct place, which would lead to a significant drop in efficiency \cite{catrysse2000}.  Instead, the lenslets are spaced slightly closer together than the pixels, which means there is an offset between lenslets and pixels at the edges of the sensor, as shown in \rfig{lenslets}.  This ensures that light is correctly focused onto each pixel when using the stock lens, and is known as Chief Ray Angle (CRA) compensation \cite{wandell2005, imx219_datasheet}.  Unfortunately, this comes at the expense of worse response at the edges of the sensor when used with optics that have a different CRA distribution across the sensor, such as microscope optics (where the CRA is nearly constant).

\fig{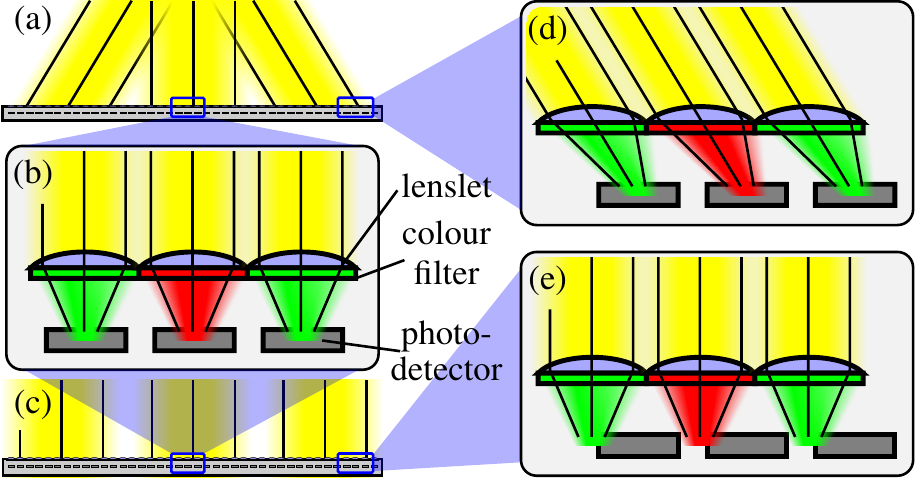}{
      In a typical webcam or mobile phone camera, light is incident at oblique angles on the edges of the sensor (a).  A very simplified representation of a pixel includes a lenslet, a colour filter, and a photodetector.  The lens focuses light onto the light-sensitive photodetector, shown in (b) for normal incidence.  Lenslets are displaced laterally at the edges of the sensor (d).  If light is incident normally across the sensor (c), this causes a drop in efficiency and a rise in crosstalk (e).
}

When using a sensor that has lenslets and CRA compensation, any change to the angle of incidence on the sensor may lead to either or both of a loss of efficiency (due to light being focused on non-light-sensitive regions of the camera) or crosstalk (due to light passing through a lenslet but hitting the wrong pixel on the other side).  Crosstalk between pixels is particularly problematic because it leads to confusion between the colour channels \cite{jian2009}; pixels are arranged in a Bayer pattern, alternating between Red, Green, and Blue filters (with twice as many green pixels), so any light displaced from one pixel to an adjacent pixel will lead to a decrease in saturation.  The Bayer pattern means that the four nearest neighbors of each red or blue pixel are all green, thus the crosstalk is strongest between green and the other two channels.

\subsection{Image processing pipeline}
The Raspberry Pi GPU firmware includes routines for handling data from the camera module.  This not only handles the data stream from the sensor (which streams raw pixel data directly to the GPU) but also converts the raw images into RGB images for display, via several denoising and correction steps.  The processed, corrected images can then either be displayed (with very low latency), compressed as JPEG or H264, or returned to the CPU for storage, further processing, or streaming over the network.  The exact details of the algorithm are proprietary, as the GPU firmware is not open.  However, the steps in the pipeline are documented, for example in the documentation of the \texttt{picamera} Python module \cite{picamera_pipeline}.  They include:
\begin{itemize}
\item Digital gain
\item Lens shading correction (compensates for vignetting on the sensor/lens combination)
\item White balance correction
\item Denoising on raw Bayer data
\item Demosaicing (conversion of Bayer-patterned image to RGB)
\item Further denoising
\item Adjusting sharpness, saturation, brightness and contrast
\item Resizing to the requested resolution
\end{itemize}

The step of most interest for flat-field correction is the ``lens shading correction'' step, which is the only one that is not applied uniformly over the whole image.  This step multiplies each pixel by a gain value, to compensate for non-uniformity in the response of the sensor.  These gain values are stored in a ``lens shading table'' with a resolution that is $\tfrac{1}{64}$ of the sensor's native resolution, and interpolated smoothly to give values for each pixel.  The pipeline for version 1 of the camera module used a fixed lens shading table that compensated for the characteristics of the OmniVision OV5647 sensor and lens.  With version 2 of the camera module, a more sophisticated adaptive lens shading algorithm is used, which is capable of adapting to different lenses (to the best of our knowledge, the details of this algorithm are proprietary).  It is now possible to supply a custom lens shading table thanks to updated \texttt{userland} libraries, which is what allows us to apply a calibration for vignetting in real time, when not using the stock lens.

\section{Measuring the sensor's response}
As discussed in various forums \cite{raspberrypi_forum_lens_shading}, calibrating a new lens/sensor combination is, in general, a difficult problem requiring careful measurements and specialist equipment.  Images must be acquired of a very uniform white target, under even illumination, and care must be taken to avoid artefacts due to stray light or dirt on the target, lens, or sensor.  The basic principle, however, is quite simple: acquire a raw image that \emph{should} be uniform and white, then calculate the required gains as the reciprocal of that image -- this will then mean that the corrected image of that target will be white.  Of course, it also means that any imperfections in the image will be transferred to future images via the lens shading table that is used in subsequent processing.

\subsection{Calibration apparatus}
In general, the response of an imaging system will depend on the optics used and the sensor.  However, there is an important special case, which is that of an imaging system that is very close to uniform, with light normally incident across the whole sensor.  As the IMX219 sensor, like the sensor in most mobile phone cameras, is only $4.6\,$mm across, this assumption is reasonable when using most good quality lenses with a focal length that is much longer than the sensor size.  This includes a great many lenses designed for traditional or digital SLR cameras (where the film or sensor is typically $10\times$ larger than our sensor) and most microscopes.  If we need only ensure that the sensor is uniformly illuminated, at normal incidence, then it is not necessary to image a perfectly uniform test target - instead we can simply illuminate the sensor with a light source that is spatially quite small, placed some distance from the sensor (to ensure uniform, approximately collimated illumination).  Such an arrangement is easy to construct using a tri-colour LED and a 3D printed collimation tube, as shown in \rfig{apparatus}.  In order to avoid issues due to PWM control of the LEDs causing stripes in the image due to the camera's rolling shutter, the light from the NeoPixel was deliberately attenuated by placing a diffuser (white paper) $8\,$mm from the LED.  This meant that the LED could be set to full brightness, without saturating the camera's sensor.  

\fig{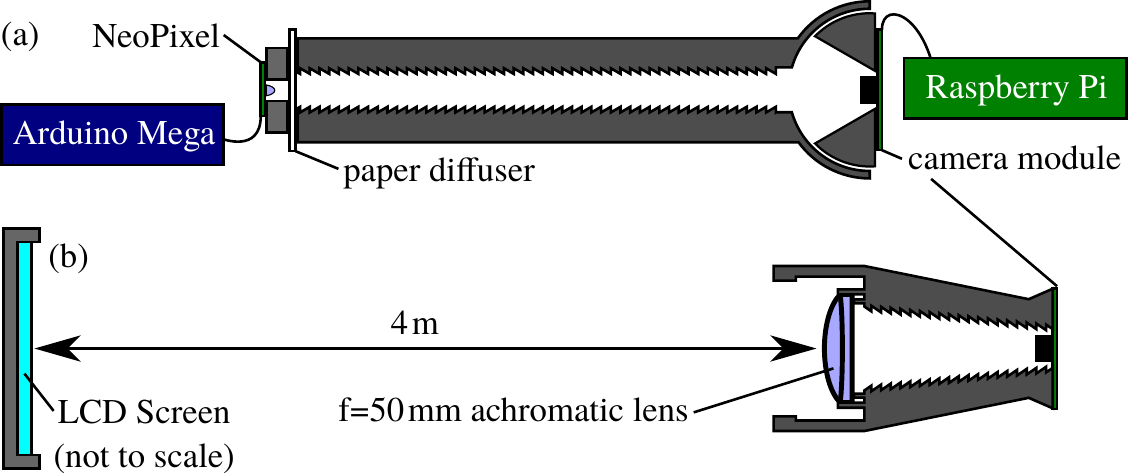}{
      (a) The calibration jig where a NeoPixel RGB LED illuminates a diffuser, and then provides uniform illumination over the camera sensor.  The camera module is able to rotate to vary the angle of incidence.  (b) A simple imaging set-up, where a computer screen is imaged using a ThorLabs AC127-050-A achromatic lens.  The plastic housing in (b) is an OpenFlexure Microscope infinity corrected optics module, without the lens present.  Printable files for both set-ups are available in the project repository \cite{cra_compensation_gitlab}.
}

This arrangement allows us to measure images taken under uniform, collimated white illumination, as well as recording individual images under red, green, and blue illumination, and indeed images with no illumination.  Python code used to acquire these images, as well as Arduino code used to control the NeoPixel LED unit, is included in the accompanying archive to this paper.

\subsection{Measurements}
Our experiment consisted of acquiring five images to calibrate the sensor, and a further image containing colour wheels at different points in the camera's frame to test our calibration.  At the start of the experiment, the sensor was run in full auto mode for a few seconds, to adjust the exposure time, gain, and white balance to produce an image that was not saturated and had neutral colour balance, under white illumination.  After this had been achieved, the settings were frozen, and the illumination was changed through different values, and images were saved with each of white, red, green, blue, and black illumination.  Finally, the test image was acquired and saved, using the same camera settings.  An example set of images is shown in \rfig{example_run}.

\fig{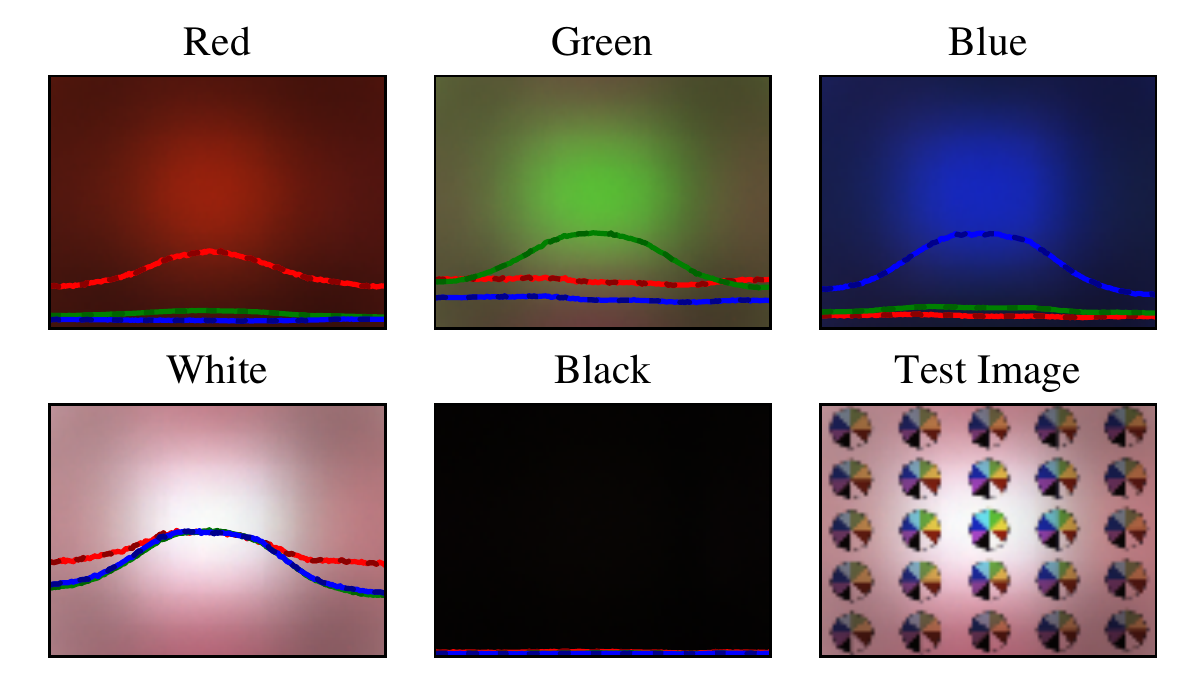}{
      A typical set of 5 images, acquired under uniform illuminations of different colours, plus a test image with multiple colours.  Note the vignetting, uneven colour response, and low saturation at the edges of the test image. Sections through the centre of each image in Red, Green, and Blue colour channels are overlaid as lines across the middle of each plot.
}

\subsection{Loading raw images}
The Raspberry Pi camera firmware is now capable of returning raw images as well as processed ones, and does so by embedding the raw data into a JPEG image file.  It is important to note that the image represented by the JPEG data has generally been through most of the pipeline described earlier, and is thus not the same as the raw image that appears later in the file.  We use a modified version of the \texttt{picamera.array} submodule to strip the raw data from the end of the JPEG file, and reconstitute it into a \texttt{numpy} array \cite{oliphant2006}.  Pixels are assigned to the appropriate colour channel, but no demosaicing or smoothing is performed.  Each pixel thus has only one non-zero value in its red, green, and blue channels.  Binning the image in blocks of $16\times 16$ pixels then yields one RGB value for each block.

A very important consideration, particularly when the light level is low, is the black level of the image.  Multiplying the sensor's output by the inverse of a nominally white image makes the assumption that the sensor's output will be zero in the absence of light.  The IMX219 chip includes non-light-sensitive pixels, which allow it to measure and correct for the black level, which varies with temperature.  The raw 10-bit image data is adjusted such that pixels read $64$ when there is no light incident on the sensor, so when reading the raw data it is important to correct for this.

\section{Analysis}
\subsection{Vignetting} \label{sec:vignetting}
The most prominent effect noticable in the raw images is vignetting; the image is significantly less bright towards the edges of the sensor.  This is not uniform across the colour channels, in part because the lenslets do not focus the light achromatically, so we expect a more diffuse focus (and thus a gentler fall-off with distance from the centre of the sensor) using red light.  When using optics that are free from vignetting, such as a lens designed for a much larger sensor, or the calibration jig in \rfig{apparatus}(a), the Raspberry Pi camera module still exhibits significant vignetting.  To substantiate our assertion that this is due to the CRA compensation of the sensor, \rfig{straight_vs_tilted} shows a set of images with normally incident light, and light incident at an angle.  The vignetting pattern shifts noticably, demonstrating that this is an angle-dependent effect.

\fig{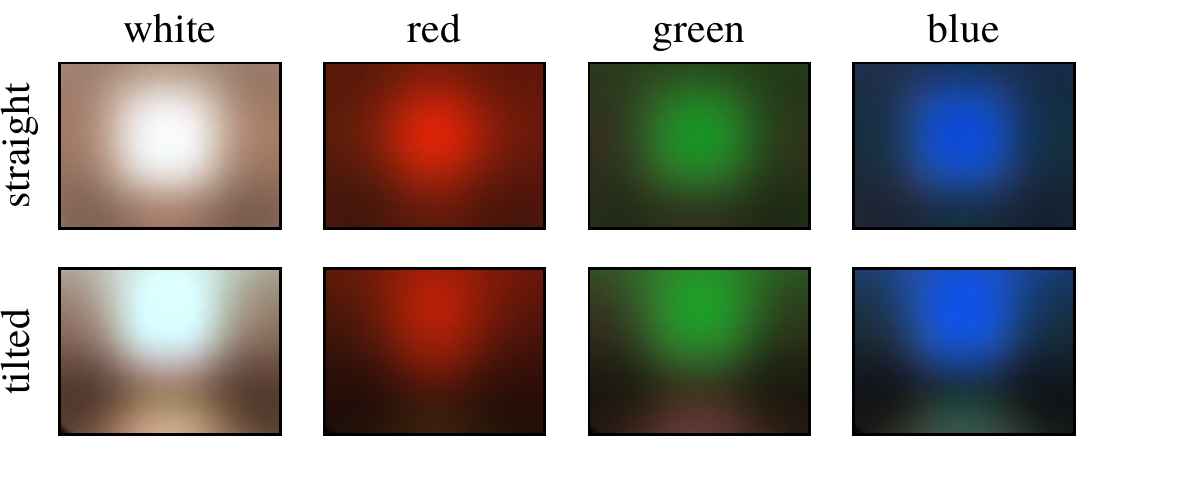}{
      Images acquired under different uniform illuminations in the calibration jig shown in \rfig{apparatus}(a). Light was either normally incident on the sensor (top row) or incident at an angle (bottom row), achieved by rotating the camera module in its cylindrical holder.  The vignetting pattern clearly shifts, demonstrating that it is largely due to the angle of incidence.  All images are normalised relative to the centre of the normally-incident white image.  Note the crosstalk where green illumination is detected by red pixels in the tilted green image.
}

A reasonable model of vignetting might be a smooth 2D image, which may be different for the different colour channels, representing the efficiency as a function of position on the sensor.  Indeed, this is how the correction is applied in the GPU processing pipeline.  Vignetting can be corrected by dividing through by a reference image to compensate for the difference in efficiency between the centre and edges of the sensor.  If we use a white image to normalise the sensor, we will correct this fall-off in other images that are similarly illuminated.  For many applications, this is sufficient - if the images collected usually have a uniform white background, for example in bright-field microscopy, correcting the vignetting restores the background to make it uniform again.  However, it is interesting to note the vignetting-corrected images under red, green, and blue illumination; there is a clear decrease in saturation towards the edges of the image as can be seen in \rfig{example_run_normalised}.  Vignetting correction as applied to a test image of colour wheels can be seen in \rfig{colour_wheels_images}

\fig{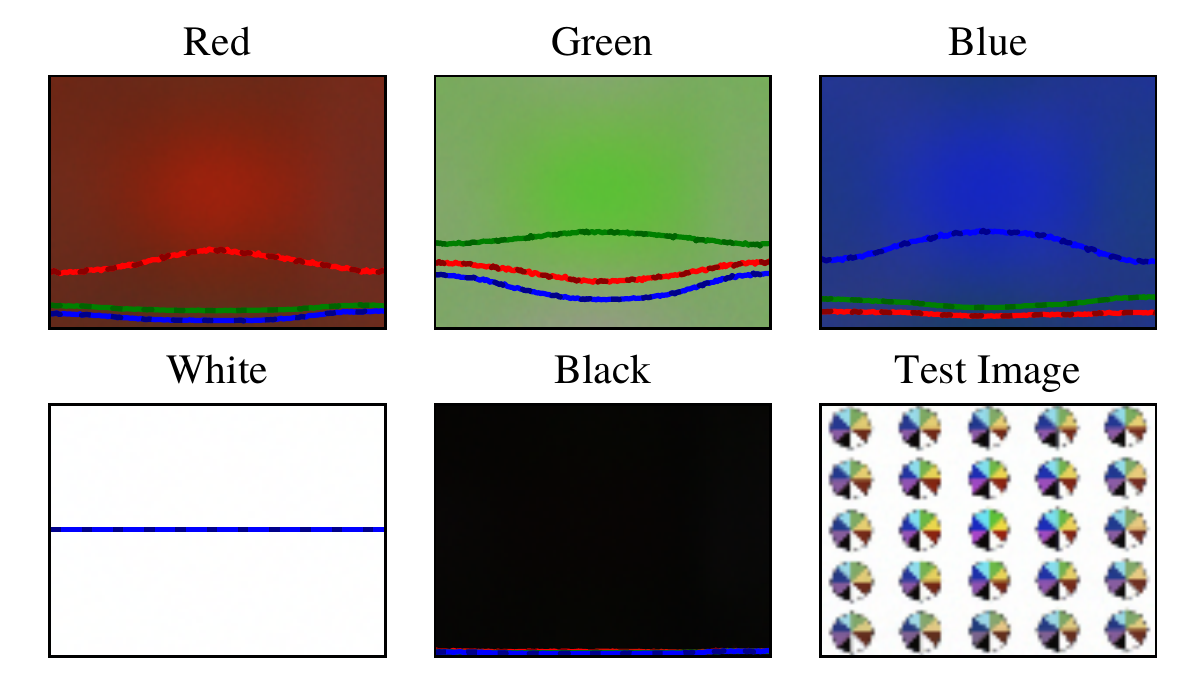}{
      The same set of images presented in \rfig{example_run}, normalised to the white image.  This is equivalent to the built-in lens shading correction, and will make the white image uniform and white by definition.  RGB components of a line section through the image centre is overlaid on each plot, as in \rfig{example_run}.
}

\fig{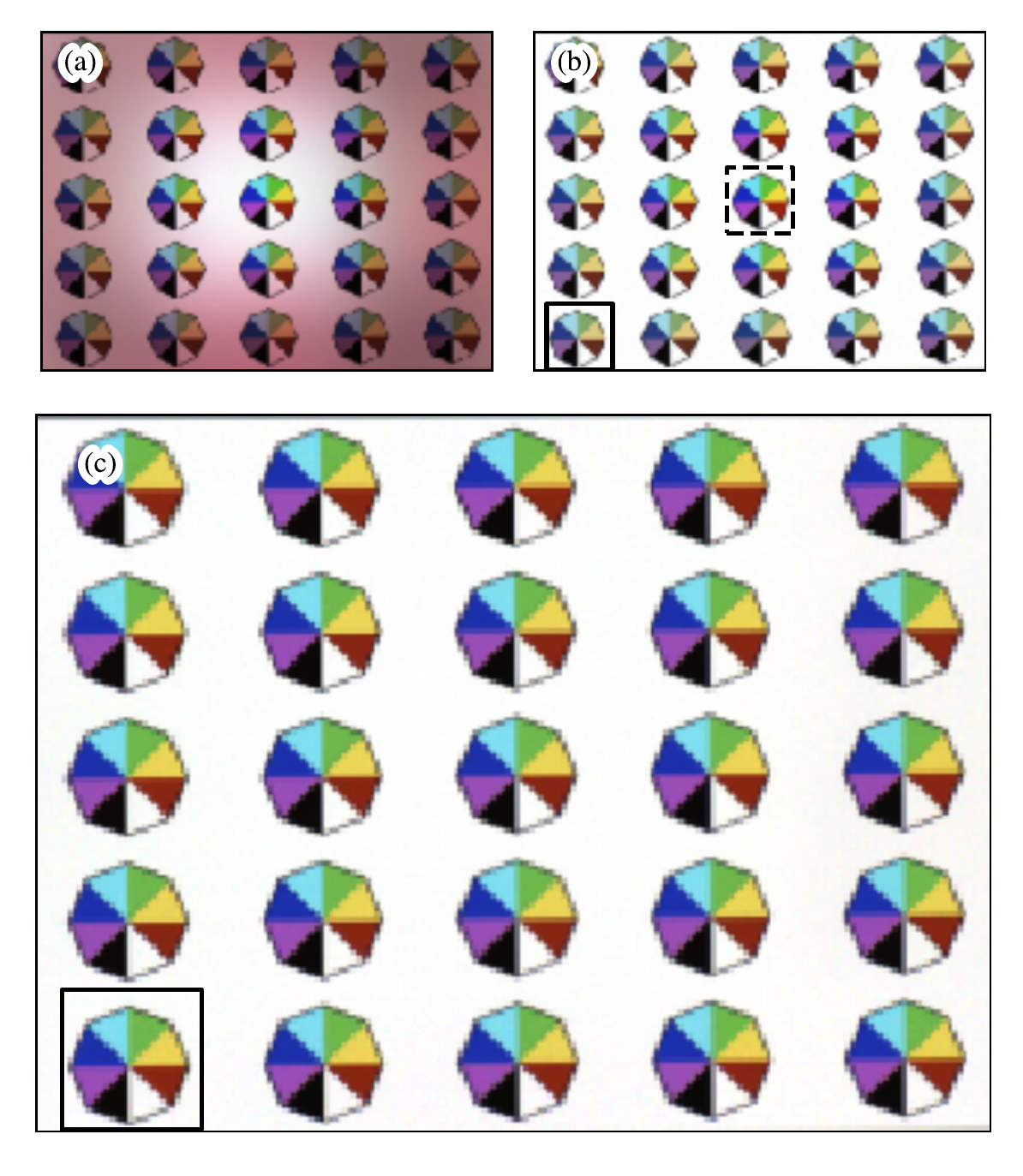}{
      Images of the colour wheel test target, with (a) only colour-balance correction $\mathrm{Cb}_{qq}$, (b) vignetting correction $\mathrm{Cb}_{qq}\mathrm{LST}_{ijq}$, and (c) colour-unmixing $\mathrm{U}_{ijpq}\mathrm{Cb}_{qq}\mathrm{LST}_{ijq}$.  The colour wheels extracted in \rfig{matrix_correction_wheels_and_polar} are outlined with black boxes.
}

\subsection{Colour crosstalk}
This decrease in saturation indicates that there is crosstalk between the colour channels - this is expected due to the lenslets not being centred on the pixels underneath at the edges of the sensor.  However, it is reasonable to suppose that this crosstalk is a linear effect and can thus be modeled and compensated for \cite{jian2009}.  While vignetting requires only one scalar parameter per channel, crosstalk requires a matrix capturing the influence of each colour channel on each of the others.  This matrix, as with the efficiency parameters for vignetting, will vary as a function of position on the sensor.  

We can split the correction of the image into a few different steps, each represented by a different matrix.  This is helpful because it means we need not acquire enough information for the entire correction each time we recalibrate.  The components we consider are:
\begin{description}
\item[Lens shading table/white image normalisation] $\mathrm{LST}_{ijk}$ has $w\times h\times 3$ degrees of freedom, i.e. it is an image.  In this analysis, we correct for vignetting before correcting for colour crosstalk.  Note that the Raspberry Pi's pipeline is usually set up with each channel normalised to have unity gain for the brightest pixel, so that the lens shading table does not affect colour balance.  Simply dividing through by a white image applies both the lens shading table and the colour balance.
\item[Colour balance matrix] $\mathrm{Cb}_{pq}$ is a diagonal $3\times 3$ matrix with three degrees of freedom, responsible for equalising the gains of the different components of the image.  Normally the gain for the green channel is fixed at unity (in order to avoid changing the overall brightness too much) and thus there are only two meaningful degrees of freedom.
\item[Colour response matrix] $\mathrm{Cx}_{pq}$ is $3\times 3$ i.e. it's a matrix that maps colour to colour, and is not spatially varying.  It should depend on the overlap of the spectra of the three illumination colours used and the camera's colour filters.  This matrix may also apply colour balance, so if $\mathrm{Cb}_{pq}$ is to be meaningful, we must define this matrix such that it does not change the overall balance of the colours.  This gives it six meaningful degrees of freedom.
\item[Spatially varying colour-unmixing matrix] $\mathrm{U}_{ijpq}$  is a $w\times h \times 3 \times 3$ array, describing leakage of one channel to another.  Depending on how this is calculated, it may include both $\mathrm{Cb}_{pq}$ and $\mathrm{Cx}_{pq}$, but it is possible to split these out.  We define $\mathrm{U}_{ijpq}$ such that it does not affect the colour balance or colour crosstalk at the brightest point of the image (generally the centre).
\end{description} 

The normalisation we've done to images in the previous section, dividing through by a white image, combines $\mathrm{LST}$ and $\mathrm{Cb}$, i.e. if we have a white image $W_{ijp}$ (where the white level is 255),
$$ \mathrm{Cb}_{pp}\mathrm{LST}_{ijp} = \frac{255}{W_{ijp}} $$

If we start with an unnormalised image $\mathrm{Image}_{ijp}$, we can colour-correct and unmix by doing:
$$\mathrm{Pure}_{ijp} = \sum_{qr}\mathrm{U}_{ijpq}\mathrm{Cx}_{qr}\mathrm{Cb}_{rr}\mathrm{LST}_{ijr}\mathrm{Image}_{ijr}$$

This will unmix the colours such that the images taken under red, green, and blue illumination become pure red, green, or blue.  However, if we just want to normalise colour response across the sensor (which has lower noise and more normal-looking colours) we simply remove $\mathrm{Cx}$:
$$\mathrm{Unmixed}_{ijp} = \sum_{q}\mathrm{U}_{ijpq}\mathrm{Cb}_{qq}\mathrm{LST}_{ijq}\mathrm{Image}_{ijq}$$
An example matrix is $\mathrm{U}_{ijpq}$ shown in \rfig{unmixing_matrix_and_snr}(a).

\fig{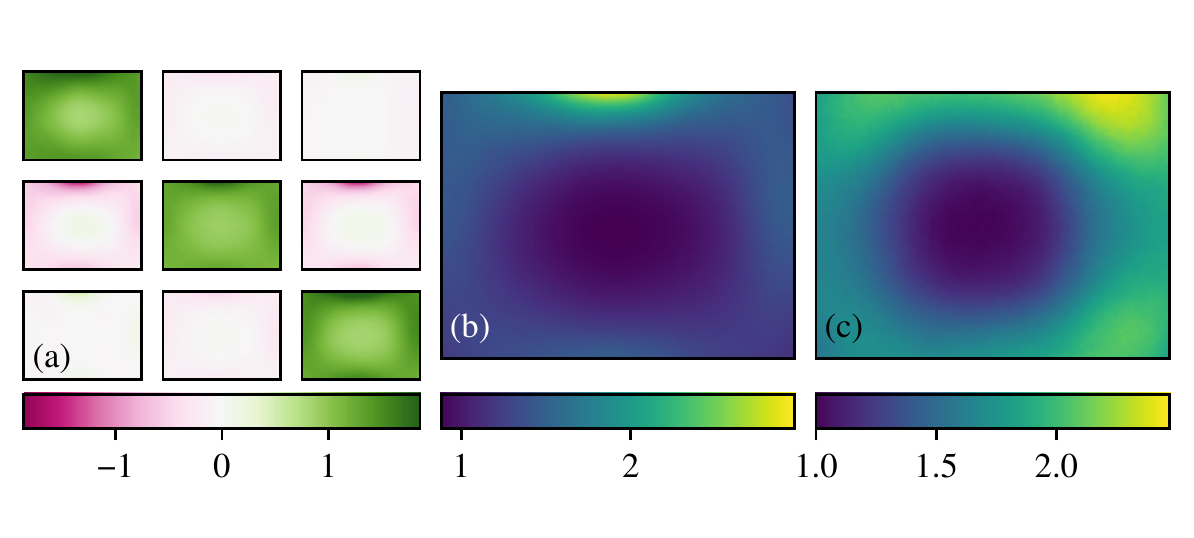}{
      (a) A representation of the $3\times 3$ spatially-varying matrix that unmixes the colour channels at the edge of the image. Crosstalk is strongest between green and the other two channels due to the spatial arrangement of the Bayer pattern, which means red and blue are never adjacent pixels. (b) The increase in noise in the image, due to the unmixing of the colour channels by the matrix in (a). This is when the colours are normalised to the centre of the image; if the colours are fully unmixed, the oversaturated image has $2.3$ times more noise again. (c) The additional noise penalty of normalising the image to correct for vignetting.
}

Both of the spatially varying matrices (i.e. the vignetting correction and the colour-unmixing matrix) should be smooth, as the effects they are correcting for ought to vary smoothly across the sensor.  Any high-frequency components in these matrices probably indicate dirt in the optical system, or some other problem.  For this reason, and to keep computational cost low, we have reduced the images in size by a factor of 16 by taking the mean of a square bin of pixels.

In order to calculate $\mathrm{U}_{ijpq}$, we must use the images acquired under red, green, and blue illumination.  In order to separate out vignetting ($\mathrm{Cb}_{kk}\mathrm{LST}_{ijk}$), we first normalise each of the single-colour images by dividing through by a white reference image.  If we stack these three images together, the result has four dimensions, with size $w\times h \times 3 \times 3$.  This matrix describes the response of each pixel to red, green, and blue light, in a $3\times 3$ matrix.  Inverting this matrix for each pixel obtains $\mathrm{Cx}_{qr}\mathrm{U}_{ijpq}$, i.e. it will convert the RGB values such that the image is uniform and fully-saturated red, green, or blue for each of the illumination values.  We use \texttt{numpy.linalg.inv} to invert the crosstalk matrix for each pixel in a Python \texttt{for} loop; this is tolerably fast for our downsampled images, despite the computational inefficiency of the interpreted \texttt{for} loop.  Using this matrix to correct the images produces fully-saturated colours, which has two drawbacks; first, this will tend to make the colours far more saturated than they really are, and second it will introduce substantial noise into the image.

A more realistic image is obtained if we normalise to the centre of the sensor, i.e. we normalise $\mathrm{U}_{ijpq}$ to be the identity in the centre of the image.  This is a simple change to make - we simply measure the colour response to the red, green, and blue LEDs, and construct a matrix that converts between the ``pure'' colours one obtains from the inverted matrix, and the actual colours measured in the centre of the sensor.  This $3\times 3$ matrix is the inverse of $\mathrm{Cx}$.  This will partially undo the correction we have applied, and results in a less extreme change to the image.  Reducing the change results in a predictable decrease in high-frequency noise in the image.  This change in noise can be simply quantified; each colour value reported is a linear combination of a number of pixel readings. This can be represented for one pixel as $c=\sum a_ip_i$ where $c$ is the output value, $a_i$ are the coefficients (which ultimately come from the matrices we have defined previously) and $p_i$ are the pixel values in the raw image.  On the assumption that the camera readout noise is independent for each pixel, the noise in the output value will be proportional to $\sqrt{\sum a_i^2}$ while the magnitude of the output value will be roughly constant, across a well-corrected image.  As the colour unmixing matrix has larger elements towards the edges, this results in the signal-to-noise ratio (SNR) getting worse towards the edges of the image.  A plot of the increase in noise due to colour unmixing ($\mathrm{U}_{ijpq}$) and vignetting correction ($\mathrm{LST}_{ijp}$) is shown in \rfig{unmixing_matrix_and_snr}(b,c).  Note that this does not include the increased noise due to colour balance (adjusting the overall brightness of the three colour channels), and it refers to the matrix that unmixes the colours to achieve uniform response over the camera rather than fully unmixing to pure red, green and blue.  The latter would increase the noise by a further factor of $2.3$ for the optical system described here.

A comparison of the different corrections is given in \rfig{matrix_correction_wheels_and_polar}.  The matrix correction is able to recover consistent colours out to the edges of the image.  While we found that the white reference image is quite specific to each optical system, it was possible to use a colour-unmixing matrix calculated from a set of calibration images acquired in a jig to unmix the colours in an image acquired using a lens imaging a computer screen.  This suggests that the correction generated using the calibration jig is applicable to most optical systems where light is normally incident across the sensor.  Clearly if the sensor is significantly tilted, or if very different optics are used, it is likely that the correction will be less good.  However, the majority of scientific applications that require removing the stock lens do have close to normal incidence across the image, and thus would benefit from this calibration.

\fig{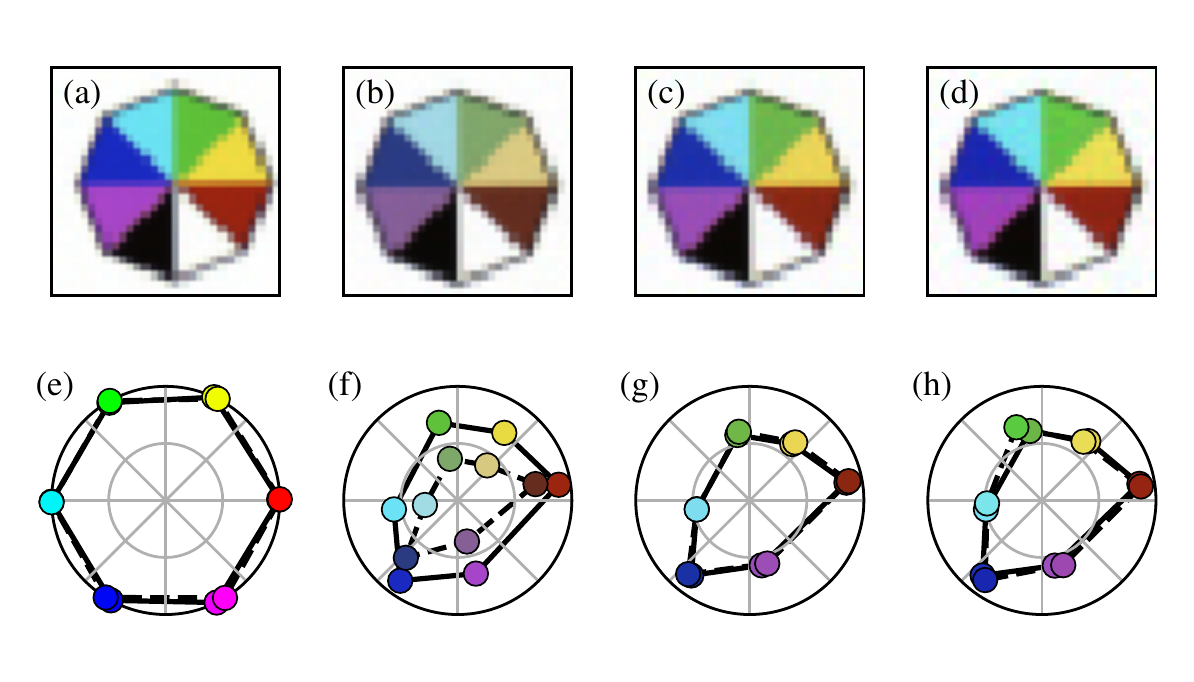}{
      Images of colour wheels extracted from the centre of a test image (a), and the corner of the same image (b-d).  Below are polar plots of the six coloured segments for the central (solid lines) and corner (dashed lines) wheels of the image (e-h). (b, f) only have vignetting correction, so the saturation is significantly lower in (b), and the two sets of colours are separated in the polar plot (f). (c, g) are unmixed using in-situ calibration, so the colours in (c) match those in (a), and the two sets of colours lie on top of each other in (g). (d, h) are unmixed using a combination of an in-situ white image and single-colour images from the calibration jig; the colours do not match as closely as in (c, g) but are much improved from (b, f). (e) is the same as (g) but unmixed to fully-saturated colours rather than matching the central response, giving oversaturated images and increased noise.  Uncropped images from which (a-c) were taken are shown in \rfig{colour_wheels_images}.
}

\section{Correcting images in experiments}
The \texttt{userland} library on the Raspberry Pi now includes bindings to the necessary MMAL functions to manipulate the ``lens shading table'' used to correct images for vignetting.  Together with the colour balance settings, this allows us to correct for vignetting ($\mathrm{Cb}_{pp}\mathrm{LST}_{ijp}$).  Unfortunately, the table of gains cannot include $\mathrm{U}_{ijpq}$ as it has only three rather than four dimensions -- colour unmixing must be done in post-processing.

The example C code provided in the \texttt{userland} library is able to take a white reference image and generate the required lens shading table to correct images. Our modified fork of the \texttt{picamera} library \cite{picamera_fork} is written in Python rather than C for ease of use, and makes it simple to include a number of improvements in the generation of the correction table.  This also greatly simplifies the procedure of dynamically adjusting the lens shading table when using the camera from a Python application.

The lens shading table has a resolution $64$ times lower than the image sensor, and is interpolated - this not only saves memory, it also enforces a certain level of smoothness in the correction (the low-resolution grid is interpolated smoothly up to the full resolution of the sensor).  We average over patches of $3\times 3$ pixels in order to reduce noise in the correction table.  

In some applications, it is relatively simple to obtain images that should be very close to uniform.  For example, in bright-field microscopy the sample can be removed, to leave what should be a uniform white image. In these situations, an in-situ calibration can be run which may have the welcome side-effect of correcting for non-uniformities elsewhere in the imaging system.  A Python routine that performs this calibration is included in the OpenFlexure Microscope software, and will be archived along with this manuscript.

Often obtaining a white image is rather simpler than obtaining the pure red, green, and blue images required for fully unmixing the colour channels.  As we have shown above, it is possible to use images acquired in a calibration jig to correct for the crosstalk between colour channels in an experiment.  This combined approach may represent a good compromise between ease of use and calibration accuracy.

\subsection{Reuse potential and adaptability}\label{h.6wkumyl0ejrh}
Our extension of the \texttt{picamera} library can be used by any hardware project that currently uses Python to acquire images from the Raspberry Pi camera.  We have also included the key routines necessary to calculate the lens shading table from a reference white image, allowing custom imaging systems to eliminate the non-uniform shading common to so many Raspberry Pi camera based projects.  Finally, the methods presented in this paper, and the accompanying Python code, allow fully colour-corrected images to be produced, if the raw images are saved and post-processed.  The same processing methods should work on JPEG images, if suitable modifications are made to the loading and saving routines.  However, we would advise caution when working with compressed images, in case the pixel values have been transformed in a nonlinear way - this would invalidate the assumption that a simple matrix inversion can compensate for colour crosstalk.

\section{Conclusion}
The lenslet array on the Raspberry Pi camera module v2 causes both vignetting and pixel crosstalk when used with optics that have close to normal incidence across the whole sensor.  Both of these effects can be calibrated out, resulting in a threefold increase in noise at the edges of the image.  Vignetting can be corrected in real time using the camera's image processing pipeline, while colour crosstalk may only be corrected after images have been transferred off the GPU.  We hope that the calibration methods and code described in this manuscript will be of use to many other projects seeking to use this extremely convenient low cost, high performance camera sensor.

\section{(4) Build Details}\label{h.l8i9vokvs0bj}

\subsection{Availability of materials and methods}\label{h.60suejv0jlzi}

The calibration jig requires a 3D printer, basic tools including hex keys, M2 and M3 screws, and elastic bands.  These are all widely available.  The electronic parts required are a Raspberry Pi computer, Raspberry Pi camera module v2, Arduino Mega and a NeoPixel LED.  The first three are widely used, and available from global suppliers such as RS Components and Farnell.  NeoPixels are available from many hobbyist electronics shops, and the generic LED module (WS2812) can be sourced more widely still.  The imaging system, used to calibrate the camera in conjunction with a display screen, uses the same camera module together with a $50\,$mm focal length achromatic lens (ThorLabs AC127-050-A) which can be obtained from ThorLabs internationally.

\subsection{Ease of build}

The calibration requires four printed components, held together with a small number of screws.  Removing the lens from the Raspberry Pi camera is the trickiest part, but it is not a difficult build.  We would estimate the calibration jig can be constructed (assuming parts are printed) in 15-30 minutes by someone who is prepared to be reasonably careful, and it does not require much previous experience.

The imaging set-up is an infinity-corrected optics module from the OpenFlexure Microscope, and is very similar to the calibration module in terms of ease of assembly.

\subsection{Dependencies, operating software, and peripherals}

The hardware components depend on a Raspberry Pi and camera module (v2) as well as an Arduino (Mega, though any model would substitute) and a single NeoPixel LED.

Calibration software is written in Python, and should be compatible with Python 2 or Python 3.  It depends on a number of relatively standard libraries (numpy, scipy, opencv-python) as well as a forked version of the picamera library (\url{https://github.com/rwb27/picamera}).  Analysis software additionally requires matplotlib, pillow, and Jupyter notebook.  Analysis does not require the picamera library and does not require to be run on a Raspberry Pi.

\subsection{Hardware documentation and files location:}

Name: GitLab

Persistent identifier: \url{https://gitlab.com/bath_open_instrumentation_group/picamera_cra_compensation/} (live repository) \url{http://dx.doi.org/TBA} (archival version)

Licence: CERN Open Hardware License

Publisher: Dr Richard Bowman

Date published: 19/7/2019

Software code repository (e.g. SourceForge, GitHub etc.)
(required)

Software is stored together with the hardware files, in the same repository.

Licence: GNU GPL v3

Date published: 19/7/2019

\section{(5) Discussion}\label{h.90jl7wm65t65}

\subsection{Conclusions}\label{h.h3fr33ylzsnh}

Conclusions, learned lessons from design iterations, learned lessons
from use cases, summary of results.

\subsection{Future Work}\label{h.neocsr410zj}

Further work pursued by the authors or collaborators; known issues;
suggestions for others to improve on the hardware design or testing,,
given what you have learned from your design iterations.

\subsection{Paper author contributions}\label{h.fy8hbipy6kwe}

RWB performed experiments, analysed the data, and wrote the manuscript.  BIV acquired the initial data and highlighted the issue in the OpenFlexure Microscope project.  JC maintains the software and performed experiments.  JS assisted with data analysis.  All authors contributed to discussions and writing of the manuscript.

\subsection{Acknowledgements}\label{h.gu3yyarx72d6}

This work would not have been possible without the underlying picamera\cite{picamera} and userland\cite{userland} libraries, which are developed by a dedicated community of volunteers.  We would also like all of the forum contributors\cite{picamera_pull_request} who have tested and commented on both the calibration code and the forked picamera library\cite{picamera_fork}.  Particular thanks is due to Dave Jones (\texttt{waveform80}) and Dave Stevenson (\texttt{6by9}), both of whom have worked with us directly as well as developing the libraries we rely on.

\subsection{Funding statement}\label{h.4u1a7tugh2om}

We would like to acknowledge financial support from EPSRC (EP/P029426/1, EP/R013969/1, EP/R011443/1), the Royal Commission for the Exhibition of 1851 (Research Fellowship for RWB), and the Royal Society (URF\textbackslash R1\textbackslash 180153, RGF\textbackslash EA\textbackslash 181034).

\subsection{Competing interests}\label{h.q1j1rznb43fl}

RWB and JS have shares in OpenFlexure Industries, which may in the future sell products based around the Raspberry Pi camera module.

\subsection{References}\label{h.6fml9tf50r5c}
\bibliographystyle{acm}
\bibliography{cra_compensation}

\section{Copyright notice}\label{h.jm5gcqv4g8x0}

Authors who publish with this journal agree to the following terms:

Authors retain copyright and grant the journal right of first
publication with the work simultaneously licensed under
a Creative Commons Attribution License that allows others to share the work with
an acknowledgement of the work's authorship and initial publication in
this journal.

Authors are able to enter into separate, additional contractual
arrangements for the non-exclusive distribution of the journal's
published version of the work (e.g., post it to an institutional
repository or publish it in a book), with an acknowledgement of its
initial publication in this journal.

By submitting this paper you agree to the terms of this Copyright
Notice, which will apply to this submission if and when it is published
by this journal.

\end{document}